# Inverted current-driven switching in Fe(Cr)/Cr/Fe(Cr) nanopillars.

M. AlHajDarwish,[1] A. Fert,[2] W.P. Pratt Jr.,[1] and J. Bass[1]
[1] Department of Physics and Astronomy, Center for Sensor Materials, Center for Fundamental Materials Research, Michigan State University, East Lansing, MI, USA 48824-2320.
[2] Unite Mixte de Physique, CNRS/THALES, Orsay, France 91404.

From both theory and experiment, scattering of minority electrons is expected to be weaker than scattering of majority electrons in both dilute Fe(Cr) alloys and at Fe(Cr)/Cr interfaces. We show that Fe(Cr)/Cr/Fe(Cr) trilayer nanopillars display a normal magnetoresistance—i.e., largest resistance at low magnetic fields and smallest at high fields, but an inverted current-driven switching—i.e., positive current flowing from the fixed to the reversing layer switches the trilayer from higher to lower resistance, and negative current switches it from lower to higher.

There is great interest in current-driven magnetic switching in nanofabricated ferromagnetic/non-magnetic/ferromagnetic (F/N/F) trilayers, both to understand the underlying physics and for device potential.[1-24] For simplest switching, one F-layer (pinned) is made much thicker (and sometimes with larger area) than the other. An applied dc current then reverses only the thinner (and sometimes smaller) F-layer (free). In all nanopillars studied so far, minority electrons were scattered more strongly than majority ones both within the F-layers and at the F/N interfaces. In such a case, a positive current flowing in the spacer layer from the thick pinned layer to the thin free one is positively polarized in the frame of the thick layer (i.e. its magnetic moment is parallel to the magnetization of the pinned layer). Large enough positive current was then found to cause the free layer's magnetic moment to rotate anti-parallel (AP) to that of the pinned layer, and reversed (negative) current caused it to rotate parallel (P). Because the scattering was the same (strongest for minority electrons) within both F-metals and at all F/N interfaces, the resistance of the trilayer was largest in the AP state at low magnetic fields H and smallest in the P state at high H, corresponding to a normal magnetoresistance (MR).[25] Sufficiently large positive current then produced a step increase in resistance and sufficiently large negative current a step decrease, which, together, we call normal switching.

If, however, majority electrons are scattered more strongly both within the pinned F-metal layer and at its F/N interface, the direction of polarization of the exiting current should reverse—i.e. positive current should be negatively spin-polarized. Published models imply that negatively-polarized positive current impinging upon the free layer should cause its moment to rotate P to that of the pinned layer,[1,2] and negative current should cause the moment of the free layer to rotate AP. If the two F layers are identical, the field-driven magnetoresistance (MR) should remain normal, with larger resistance in the low field AP-state than in the high field P-state.[25] However, the current-driven switching should 'invert', in that positive current should drive the system from the higher resistance AP state to the lower resistance P state, and negative current the opposite.

Together, theory and experiment[26-30] indicate that both a dilute Fe(Cr) alloy and an Fe(Cr)/Cr interface should scatter majority electrons more strongly. In this paper we show that an Fe(Cr)/Cr/Fe(Cr) trilayer displays the two behaviors described in the previous paragraph, namely a normal MR—resistance smallest in the high field P-state, but an inverted current-driven switching—positive current drives the system to the lower resistance P state and negative to the higher resistance AP state

Magnetic nanopillars of approximately elliptical shape and dimensions ~ 70 nm x 130 nm were prepared by triode sputtering onto Si substrates.[24] The Fe(Cr) alloy contained ~ 5 at.% Cr. The multilayers consisted of a thick Cu lower contact, a 30 nm pinned Fe(Cr) layer, a 6 nm thick Cr layer, a 3.5 nm thick 'free' Fe(Cr) layer, and a thick Au top contact. To minimize dipolar coupling between the Fe(Cr) layers, the sample was ion-milled through part of the Cr layer, so as to leave the bottom (pinned) Fe(Cr) layer wide. With this geometry, the wide layer 'switches' upon application of a relatively small magnetic field, but does not switch upon application of a current large enough to switch the patterned top (free) Fe(Cr) layer. Differential resistances, dV/dI, were measured with four probes and lock-in detection, adding an ac current of ~ 20 µA at 8 kHz to the dc current I. H is directed along the easy axis of the nanopillar.

As a comparison standard for our Fe(Cr)/Cr/Fe(Cr) data, Fig. 1 shows previously published data [24] for dV/dI vs I at H = 50 Oe and dV/dI vs H at I = 0 for nanopillars of Py/Cu/Py (Py = Permalloy = $Ni_{84}Fe_{16}$), where positive current was defined the same as in the present paper. The dV/dI vs I data are shown full size, and the dV/dI vs H (i.e., MR) data as insets in the top center. The upper curves are for room temperature (295K) and the lower ones for 4.2K. Both the MR and dV/dI vs I curves for Py/Cu/Py are 'normal'—i.e. dV/dI is smallest in the high field P state, and positive current switches the sample from its low resistance (P) state to its high resistance (AP) state, and vice versa for negative current. In the MR curves at both temperatures, the transitions from the P to AP states occur only after the magnetic field passes through zero, consistent with weak magnetic coupling.

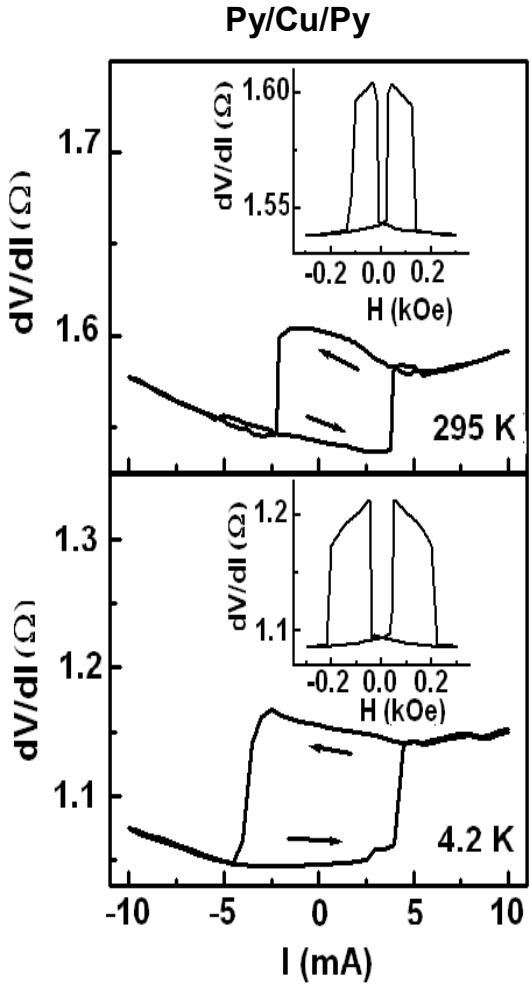
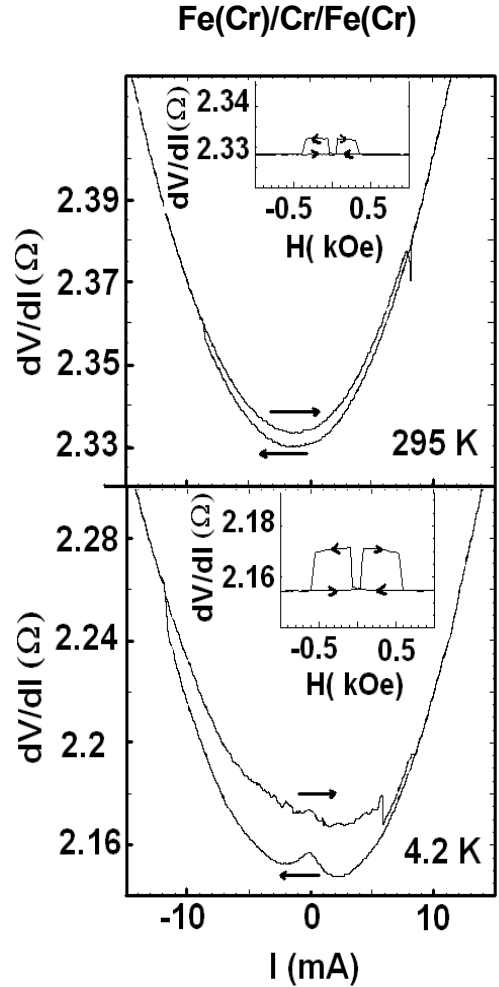

Fig. 1. Py/Cu/Py data at 295K and 4.2K showing normal switching for dV/dI vs I at H = 50 Oe (main figures) and also for dV/dI vs H at I = 0 (insets). (From Urazhdin et al.[24]).

Fig. 2. Fe(Cr)/Cr/Fe(Cr) data at 295K and 4.2K showing inverted switching for dV/dI vs I at H = 0 (main figures) but normal dV/dI vs H at I = 0 (insets).

Fig. 2 shows our new data for Fe(Cr)/Cr/Fe(Cr) trilayers. The changes in dV/dI vs I or H for Fe(Cr)/Cr/Fe(Cr) are considerably smaller than those for Py/Cu/Py, but still visible. As expected, the MR curves at I = 0 are still 'normal'—i.e., smallest in the P state. And, similarly to Py/Cu/Py, the MR transitions from the P to AP states appear only after the magnetic field passes through zero. In contrast to the similar MR curves in Figs. 1 and 2, the dV/dI vs I curves for the Fe(Cr)/Cr/Fe(Cr) samples behave oppositely to those for Py/Cu/Py. Positive I switches the Fe(Cr) free layer from the high resistance AP to the low resistance P state, and negative I switches it from the P to AP state. The dV/dI vs I curves for Fe(Cr)/Cr/Fe(Cr) are 'inverted' from the 'normal' behavior for Py/Cu/Py. In all cases in both Fig. 1 and Fig. 2, the agreement between the minimum and maximum values of dV/dI, together with almost single step switching (e.g., at –11 and + 6 ma at 4.2K), show that the current-driven switching is complete. Normal MRs and inverted current-driven switching were also seen in other Fe(Cr)/Cr/Fe(Cr) samples.

We also checked that similar results were obtained for the MR with I ≠ 0 and for dV/dI vs I for H ≠ 0. Fig. 3 shows how dV/dI vs H and dV/dI vs I change at 4.2K for the same Fe(Cr)/Cr/Fe(Cr) trilayer when we vary H at fixed I (Fig. 3a) and I at fixed H (Fig. 3b). In part because the jumps in dV/dI are small, the switching is irregular and often partial. But all dV/dI vs H switching seen is normal, and all dV/dI vs I switching seen (both hysteretic and reversible) is inverted. Asymmetry in H is presumably due to a combination of the self-Oersted field and sample shape asymmetry.

To summarize, we have shown that Fe(Cr)/Cr/Fe(Cr) trilayers give normal MR—smallest dV/dI vs H at high fields when the magnetizations of the two Fe(Cr) layers are aligned parallel (P) to each other, but inverted dV/dI vs I—large positive current causes switching from the AP to the parallel (P) state and large negative current causes switching from the P to the AP state.

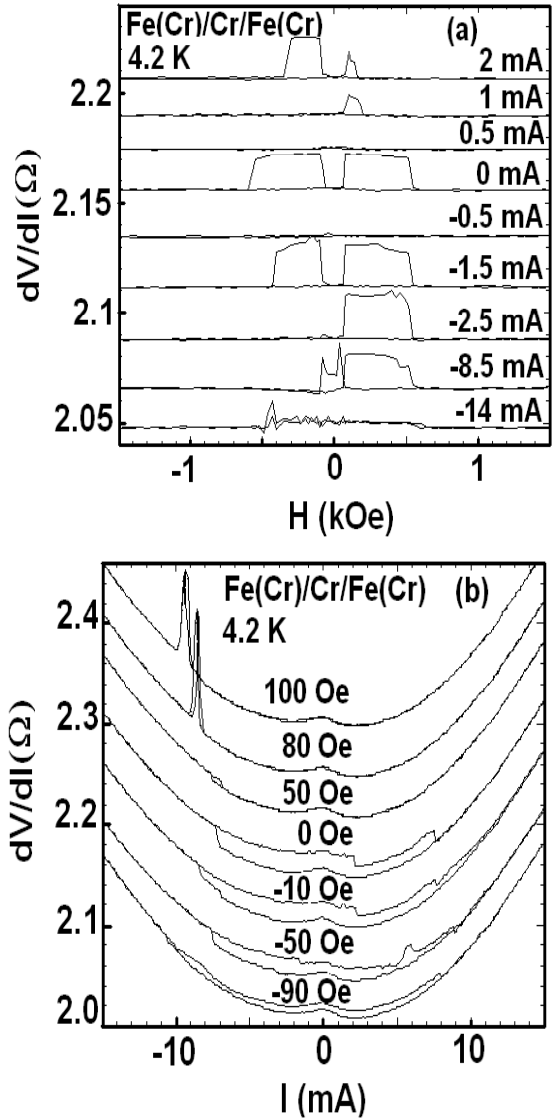

Fig. 3. Fe(Cr)/Cr/Fe(Cr) data at 4.2K showing: (a) normal MR for dV/dI vs H at various I, but (b) inverted switching for dV/dI vs I at various H. Curves in (a) for I ≠ 0 and in (b) for H ≠ 0 are shifted vertically for clarity.

Acknowledgments: The authors thank S. Urazhdin for helpful suggestions and for permission to use his published Py/Cu/Py data in this paper. This research was supported in part by the MSU CFMR, CSM, the Keck Microfabrication Facility, NSF grants DMR 02-02476, 98-09688, NSF-EU collaborative grant 00-98803, and Seagate Technology.

[1] J. Slonczewski, J. Magn. Magn. Mater. **159**, L1 (1996); J. Magn. Magn. Mat. **247**, 324 (2002).
[2] L. Berger, Phys. Rev. **B54**, 9353 (1996); J. Appl. Phys. **89**, 5521 (2001).
[3] M. Tsoi, A.G.M. Jansen, J. Bass, W.-C. Chiang, V. Tsoi, and P.W. Wyder, Phys. Rev. Lett. **80**, 4281 (1998); ibid, **81** 493 (E) (1998).
[4] M. Tsoi, A.G.M. Jansen, J> Bass, W.-C. Chiang, V. Tsoi, and P.W. Wyder, Nature **406**, 46 (2000).
[5] E.B. Myers, D.C. Ralph, J.A. Katine, R.N. Louie, and R.A. Buhrman, Science **285**, 867 (1999).
[6] Ya. B. Bazaliy, B.A. Jones and S.C. Zhang, J. Appl. Phys. **89**, 6793 (2001).
[7] A. Brataas, Y. V. Nazarov, and G.E.W. Bauer, Phys. Rev. Lett. **84**, 2481 (2000).
[8] X. Waintal and P.W. Brouwer, Phys. Rev. **B63**, 220407 (2001).
[9] C. Heide, Phys. Rev. **B65**, 054401 (2002).
[10] M.D. Stiles and A. Zangwill, Phys. Rev. **B66**, 0114407 (2002).
[11] S. Zhang, P.M. Levy, and A. Fert, Phys. Rev. Lett. **88**, 236601 (2002).
[12] J.-E. Wegrowe, Phys. Rev. **B62**, 1067 (2000).
[13] J.A. Katine, F.J. Albert, R.A. Buhrman, E.B. Myers, and D.C. Ralph, et al., Phys. Rev. Lett. **84**, 3149 (2000).
[14] F.J. Albert, J.A. Katine, R.A. Buhrman, and D.C. Ralph, . Appl. Phys. Lett. **77**, 3809 (2000).
[15] F.B. Mancoff and S.E. Russek, IEEE Trans. Magn. **38**, 2853 (2002).
[16] J.Z. Sun, D.J. Monsma, D.W. Abraham, M.J. Rooks, and R.H. Koch, Appl. Phys. Lett. **81**, 2202 (2002).
[17] E.B. Myers, F.J. Albert, J.C. Sankey, E. Bonet, R.A. Buhrman, and D.C. Ralph, Phys. Rev. Lett. **89**, 196801 (2002).
[18] F.J. Albert, N.C. Emley, E.B. Myers, D.C. Ralph, and R.A. Buhrman, Phys. Rev. Lett. **89**, 226802 (2002).
[19] Y. Ji, C.L. Chien, and M.D. Stiles, Phys. Rev. Lett. **90**, 106601 (2003).
[20] W.H. Rippard, M.R. Pufall, and T.J. Silva, Appl. Phys. Lett. **82**, 1260 (2003).
[21] J. Grollier, V. Cros, A. Hamzic, J.M. George, H. Jaffres, A. Fert, G. Faini, J. Ben Youssef, and H. Le Gall, Appl. Phys. Lett **78**, 3663 (2001).
[22] M.R. Pufall, W.H. Rippard, and T.J. Silva, Appl. Phys. Lett. **83**, 323 (2003).
[23] B. Oezyilmaz, A.D. Kent, D. Monsma, J.Z. Sun, M.J. Rooks, and R.H. Koch, Phys. Rev. Lett. **91**, 067203 (2003).
[24] S. Urazhdin, N.O. Birge, W.P. Pratt Jr., and J. Bass, Phys. Rev. Lett. **91**, 146803; S. Urazhdin, H. Kurt, W.P. Pratt Jr., and J. Bass, Appl. Phys. Lett. **83**, 114 (2003).
[25] J. Bass and W.P. Pratt Jr., J. Magn. Magn. Mat. **200**, 274 (1999).
[26] C. Vouille, A. Barthelemy, F. Elokan Mpondo, A. Fert, P.A. Schroeder, S.Y. Hsu, A. Reilly, and R. Loloee, Phys. Rev. **B60**, 6710 (1999).
[27] I. Mertig, R. Zeller, and P.H. Dederichs, Phys. Rev. **B47**, 16178 (1993).
[28] I.A. Campbell and A. Fert, Ferromagnetic Materials, **3**, E.P. Wolforth, Ed., North-Holland, 1982, Pg. 747.
[29] M. D. Stiles and D.R. Penn, Phys. Rev. **B61**, 3200 (2000).
[30] K. Xia, P.J. Kelly, G.E.W. Bauer, I. Turek, J. Kudrnovsky, and V. Drchal, Phys. Rev. **B63**, 064407 (2001).